Technical Report

# An incomplete taxonomy of self–assigned color specialties

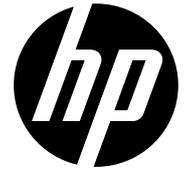

Dr. Ján Morovič, Senior Color Scientist, HP Inc.

Every discipline in science or professional practice can be sub-divided into specialties and subspecialties. E.g., in medicine a doctor could indicate that their specialty is ophthalmology and their sub-specialty the retina, or in chemistry a specialty could be analytical chemistry and a sub-specialty spectroscopy. But, what are the specialties and sub-specialties of the color discipline? The present report shares the anonymized results of a 152–participant, online survey conducted between 23$^{rd}$ May and 8$^{th}$ June 2022 that suggests 11 top–level color specialties.

## Context

Understanding what aspect of a discipline its practitioner is specialized in allows for an efficient connection between challenges that need to be solved and those who are best placed to solve them. In an increasingly distributed workforce, having an effective taxonomy of a discipline's specialties and subspecialties enables flexibility and an effective use of an organization's talent. It also facilitates the identification of present or future skill gaps.

While specialties are well understood in established disciplines, the color discipline today lacks a skills taxonomy that reflects the breadth of this multi-disciplinary field. The present report does not pretend to provide the answer to this question, but only to offer a response that may contribute to a future, unified and comprehensive taxonomy.

## Approach and limitations

The process that has led to the incomplete taxonomy of color specialties and subspecialties shared below started with the following instructions and questions shared with my contacts within the color discipline at HP and with my LinkedIn contacts:

> "Hello:
>
> To understand the variety of aspects of the broader color discipline, we would like to ask for your help with building a taxonomy of color specialties and sub-specialties. To illustrate the concept, we can look to other domains. E.g., in medicine a doctor could indicate that their specialty is ophthalmology and their sub-specialty the retina, or in chemistry a specialty could be analytical chemistry and a sub-specialty spectroscopy.
>
> If you have multiple color specialties or sub-specialties, please, select the ones where you have most expertise.
>
> This survey will close on 31st May 2022 and all answers will be anonymized. If you have any questions or comments about this survey, please, get in touch with jan.morovic@hp.com
>
> Thank you very much for your time and support. Please proceed with the survey now by clicking on the Start button below."

These instructions were followed by two questions (and the option to request a report following the conclusion of the survey:

*"What are your color specialties and sub-specialties? Please provide up to three in your own words and state them as "specialty / sub-specialty".*

*What color specialties or sub-specialties do you think will most be needed in the future?"*

The 152 responses collected between 23rd May and 8th June 2022 therefore represent my contacts (and those to whom my contacts forwarded this survey), instead of being a random sample of all practitioners of the color discipline. This sampling bias and incompleteness are the first major limitations of the present exercise. Next, the responses obtained in this way were structures by me and are therefore a consequence of my own view of the color discipline and biased by the specialties that I myself practice and am most exposed to. This subjectivity is the second major limitation. Finally, the questions were deliberately left very vague and open, which is both a limitation and a strength. The choice to not provide example answers or some initial set of color specialties was deliberate, so as to encourage each participant to describe their color specialty and sub-specialties in their own words. This choice was made to ensure that the resulting taxonomy may have the greatest chance of all color discipline practitioners finding themselves well represented by it.

In summary, the following can be thought of as my view of the color discipline based on responses of 152 of my network of direct and indirect contacts, expressed in the way in which each one of them thinks about their own experise. I therefore expect it to be incomplete (with some gaps being clear to me – e.g., the omission of color philosophy – while I may not even be aware of others), skewed towards color science and color reproduction, where my own specialties lie, and more heterogeneous than would have been the case had the questions been more guided up front.

Nonetheless, even this incomplete view shows the breadth of the color discipline and its inter-disciplinary nature very clearly and I hope that it will contribute to a more comprehensive taxonomy and one less constrained by a single practitioner's judgment calls in the future.

## Participants

The 152 participants responded to the survey from 35 countries with frequencies as shown in the following map where the color coding corresponds to the number of participant per country.

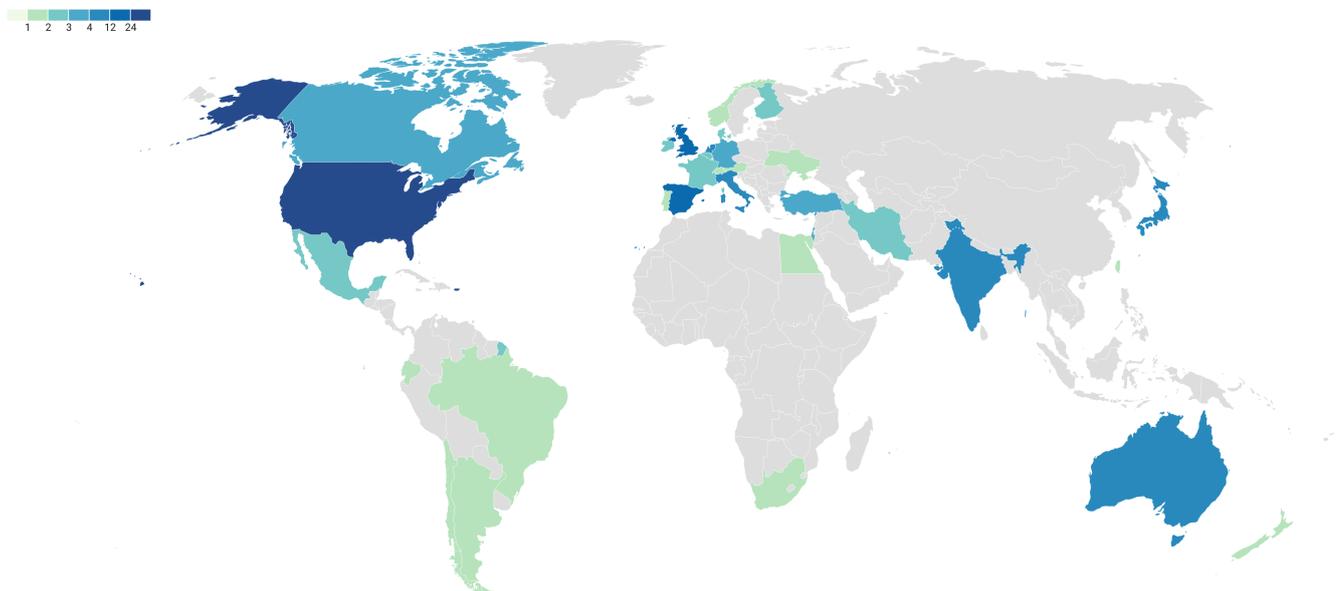

## Response structuring

The 232 distinct, raw responses to the first question of the survey were used to build the color specialty map that will be shown below. However, since it was, deliberately, posed in an unconstrained way and since the instructions invited participants to describe their own specialties in their own words, the following is very much not a direct reporting of the survey results, but an expression of the structure that I saw when looking at them. The individual items shown in the taxonomy below are, however, the terms used by survey participants, with two modifications made by me. In some cases I combined multiple answers into a single entry (e.g., "Architecture / the built environment"). In others I changed specialty-subspecialty choices (e.g., where someone would have said that their specialty is "Color appearance models", I would have inserted this as a sub-sub-specialty under "Color science" as the specialty and "Color appearance" as the sub-specialty). In other words, all entries are direct or combined participant responses, but their organization into what is considered as a specialty or sub-specialty are my choices.

# Results

The following are a series of views of the color specialty map at different levels of detail. Wherever a number in brackets is shown at the end of an entry, it indicates how many times a respondent to the survey has self-described their specialty as such (e.g., "Color management (32)" means that 32 respondents indicated that their specialty is color management). For views of the taxonomy where not all levels are included, numbers are cumulative for a specialty's sub-specialties. Where no numbers in brackets are shown, only a single responded provided the specialty shown.

## Top-level color specialty map

For the highest level, the following view omits all levels of sub-specialty and only lists the specialties under which all of the respondents' answers were grouped:

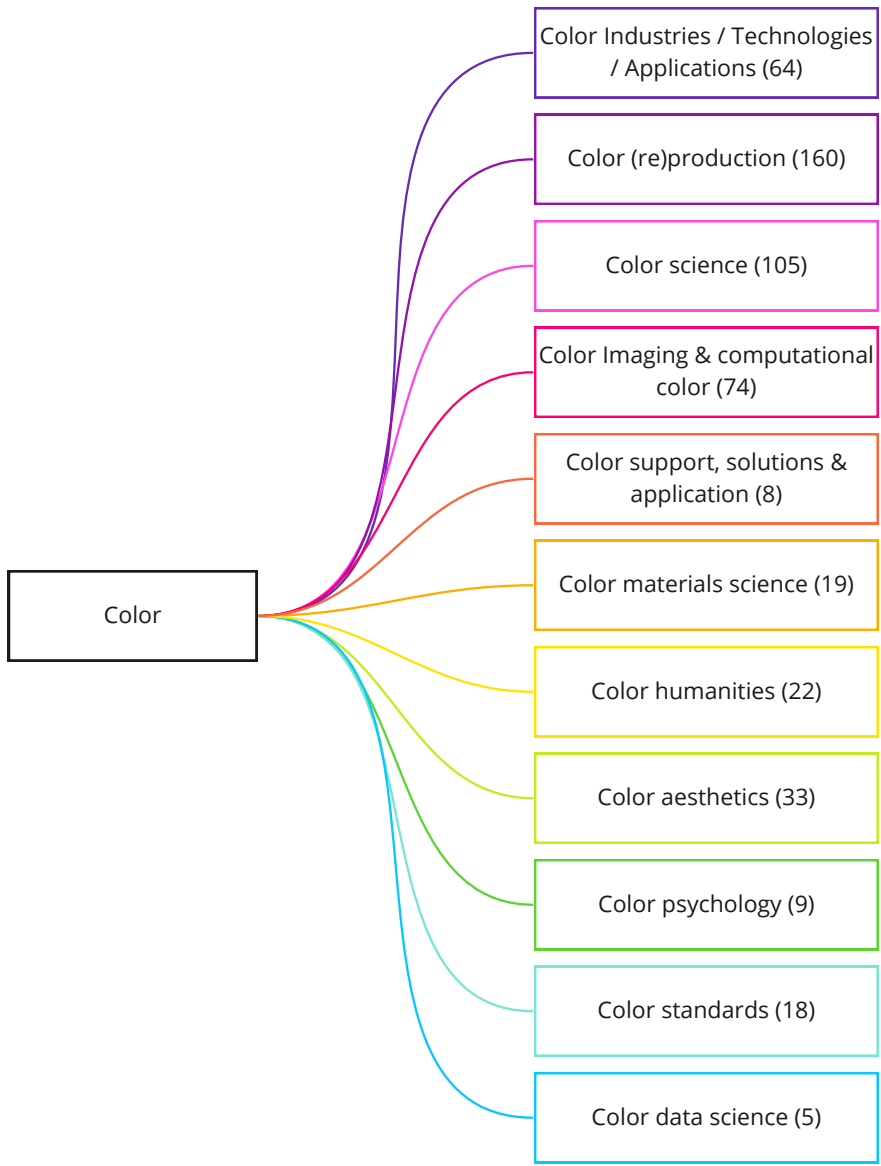

# Two-level color specialty map

For the next level of detail, the following view also shows first-level sub-specialties for each of the above specialties:

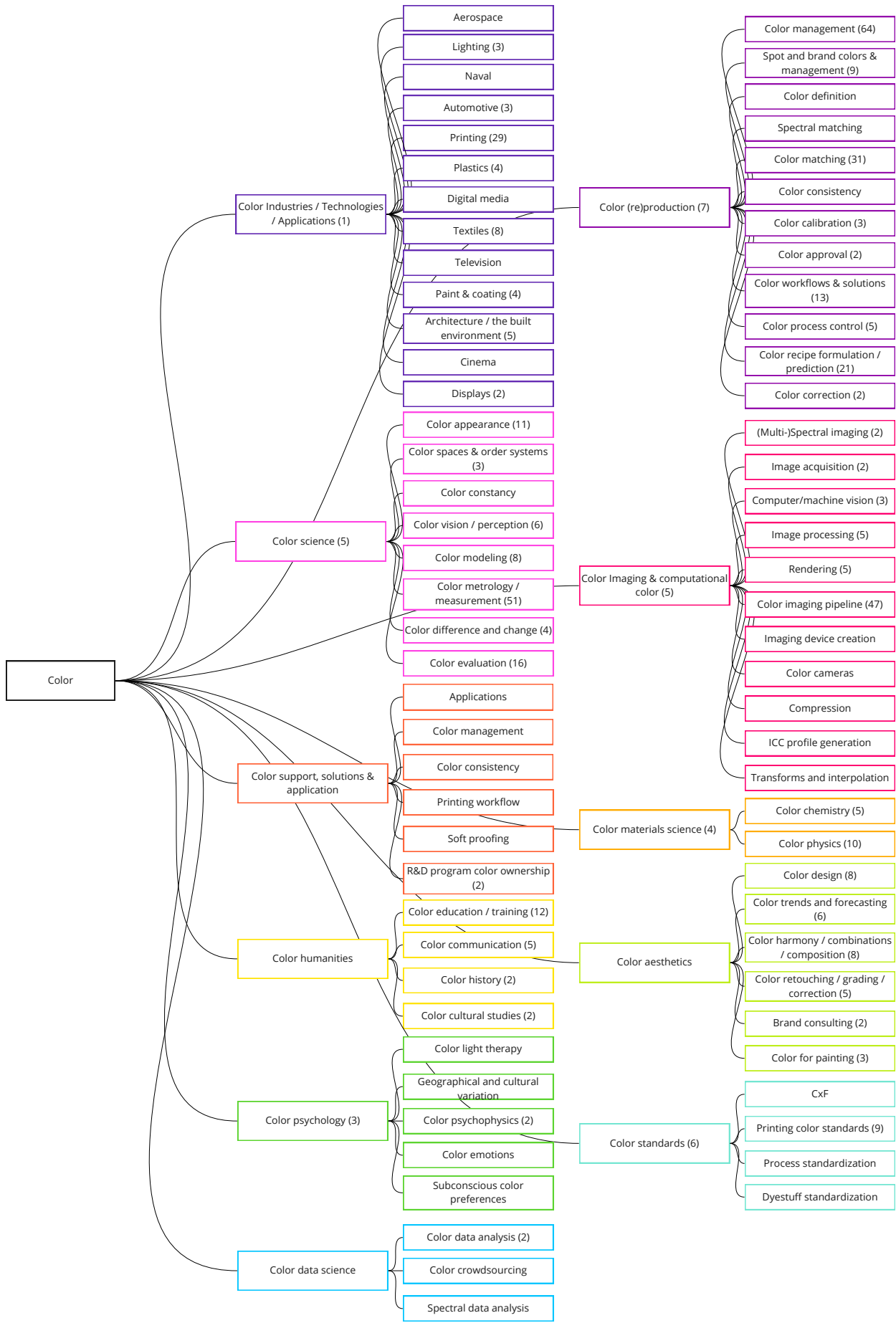

## Full color specialty map

Finally, a full map of all color specialties with up to four levels of sub-specialties can be seen in Appendices A and B at the end of this report. The following just illustrates its overall structure:

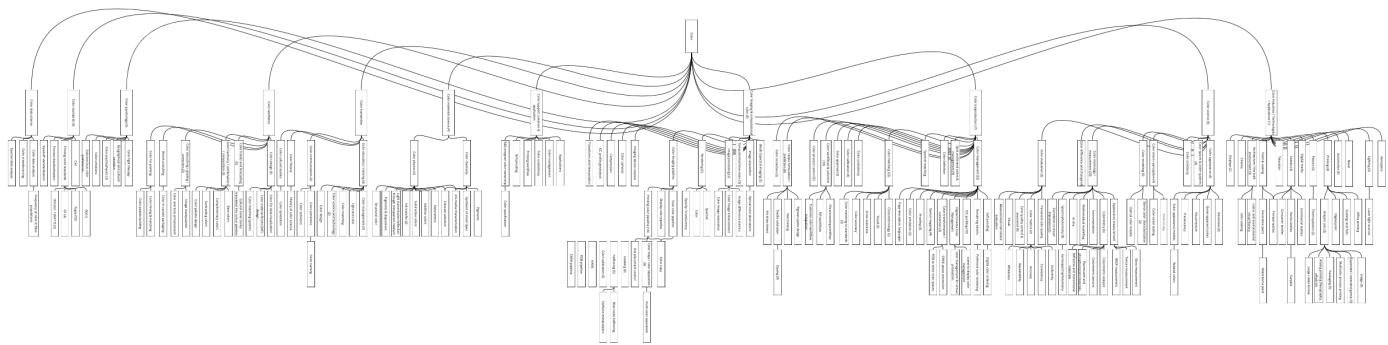

The latest version of this report can be found here: http://arxiv.org/abs/2206.08779.

For any questions, comments, or to add your specialty and sub-specialties, if they are not already shown here, please, email jan.morovic@hp.com.

# Appendix 1: Full color specialty map

# Appendix 2: Full color sub-specialty maps

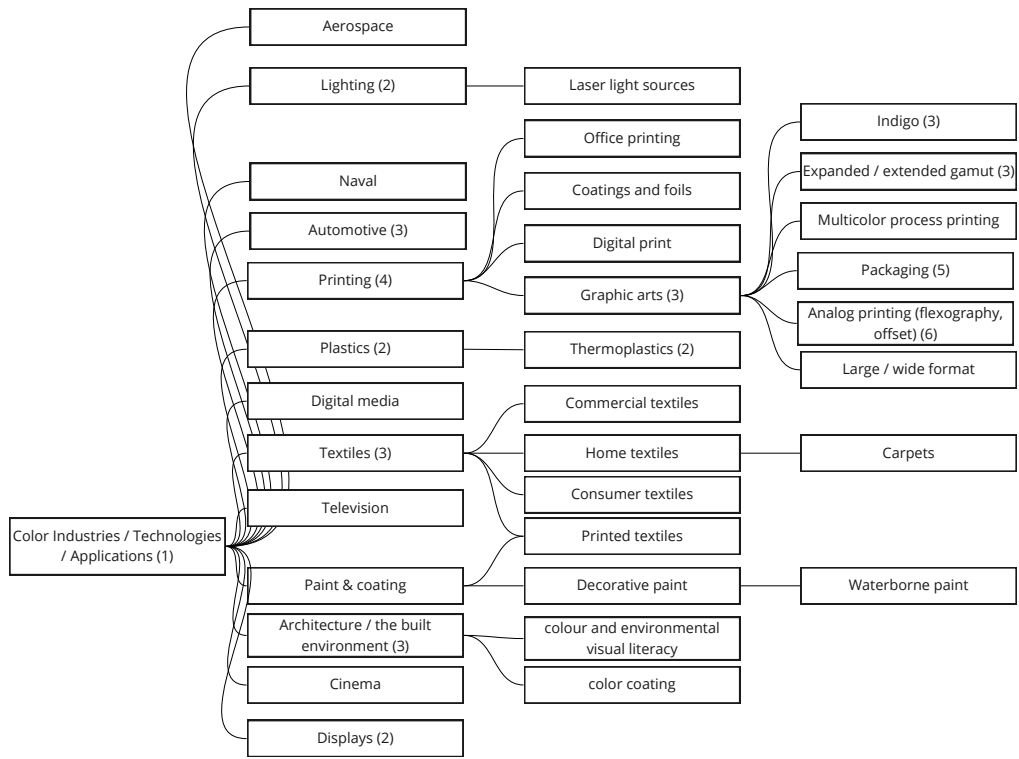

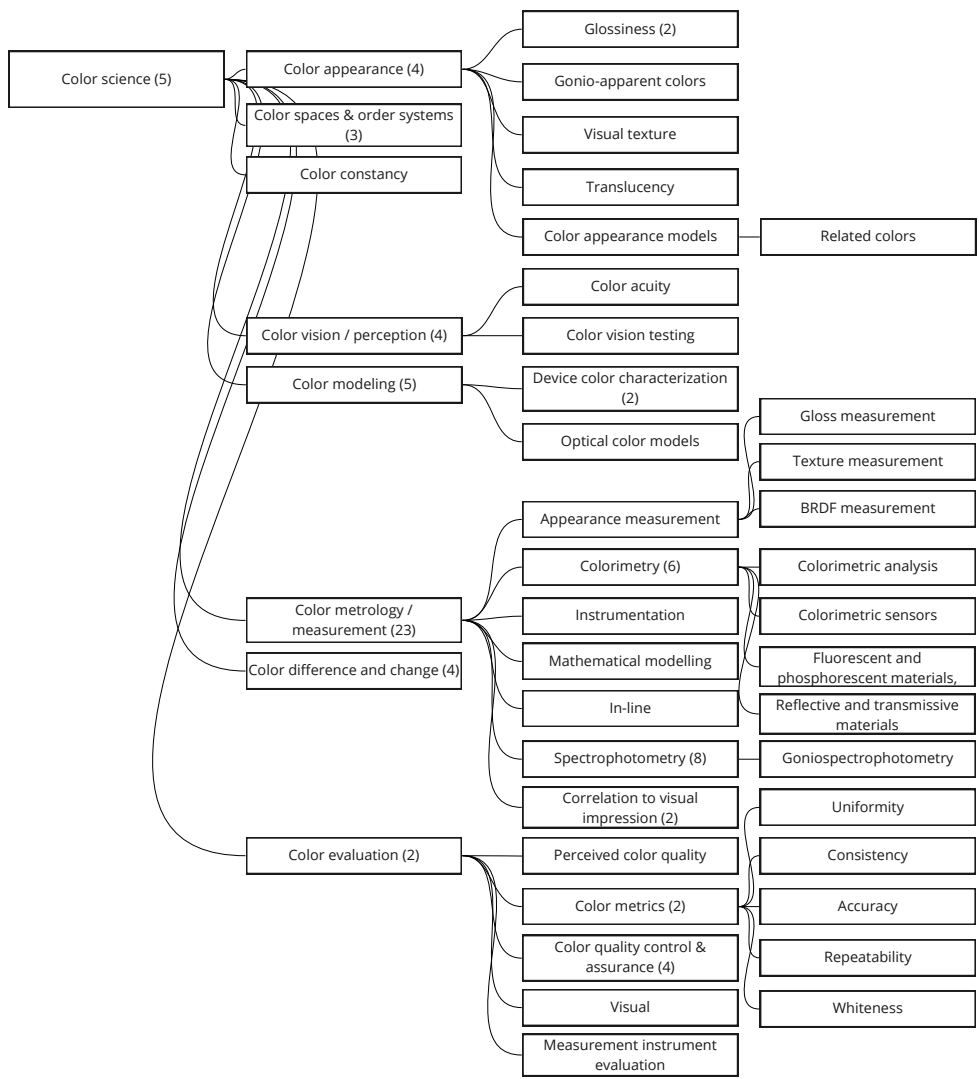

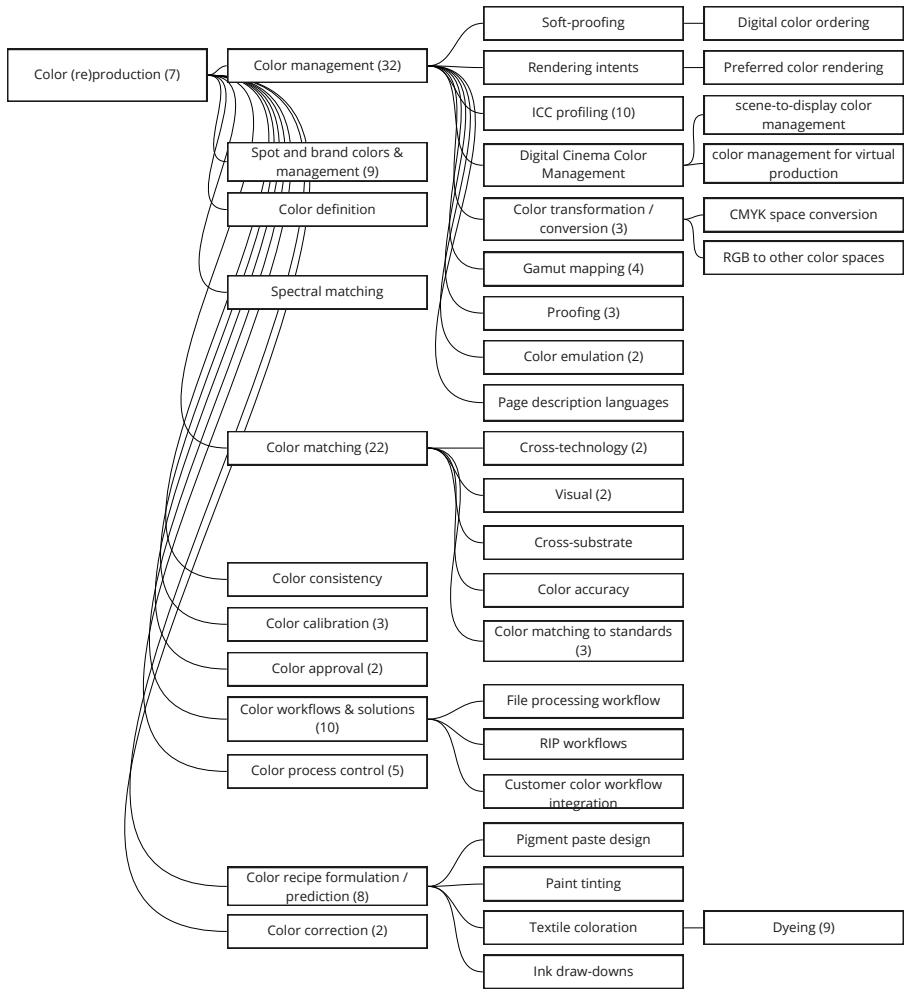
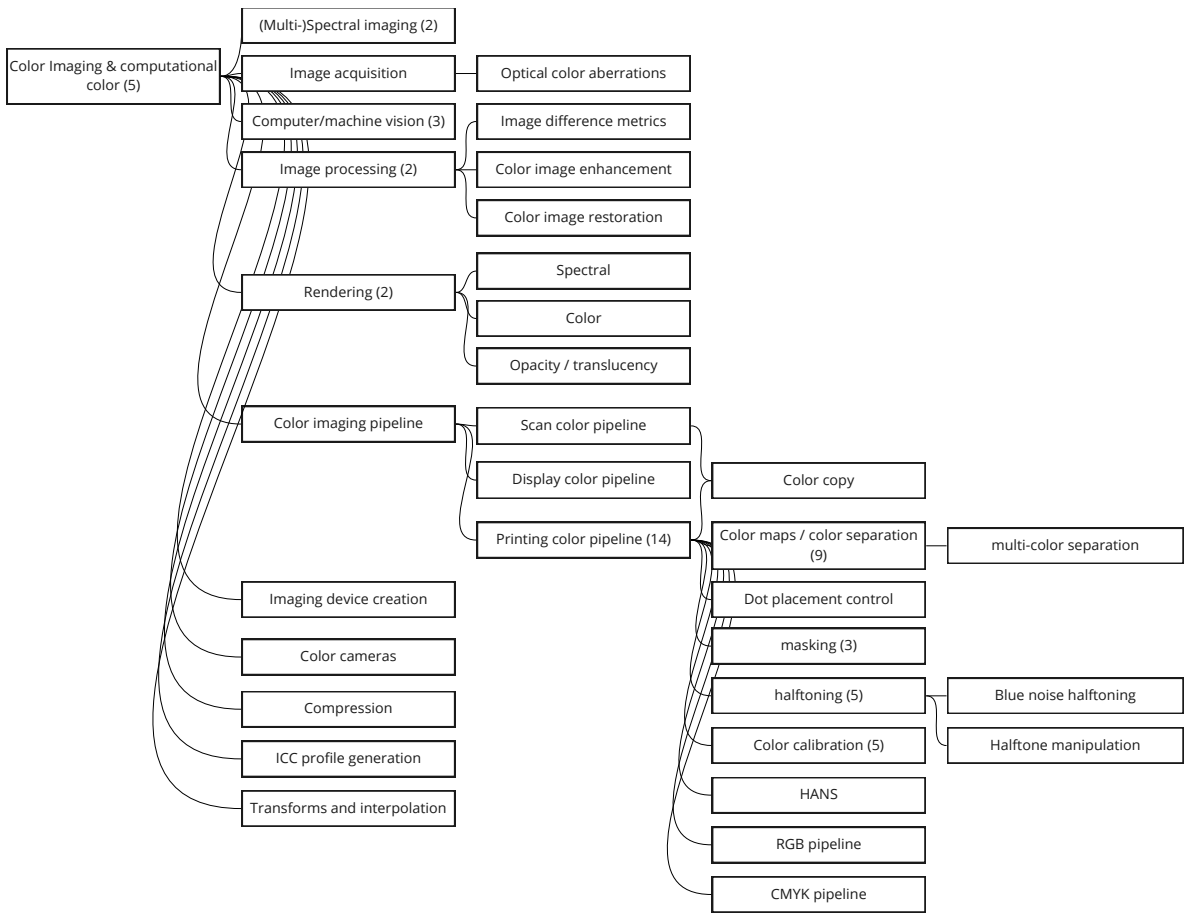

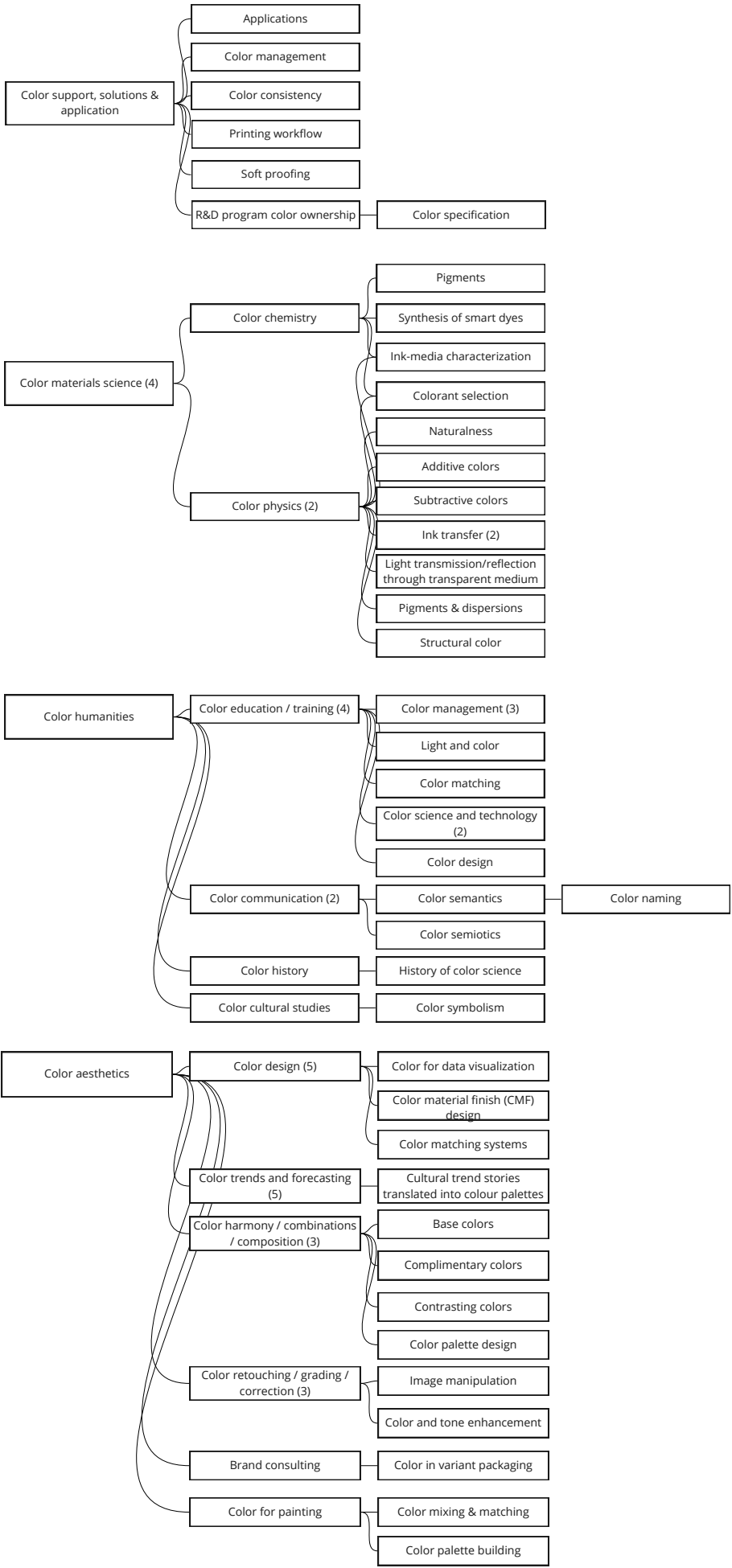

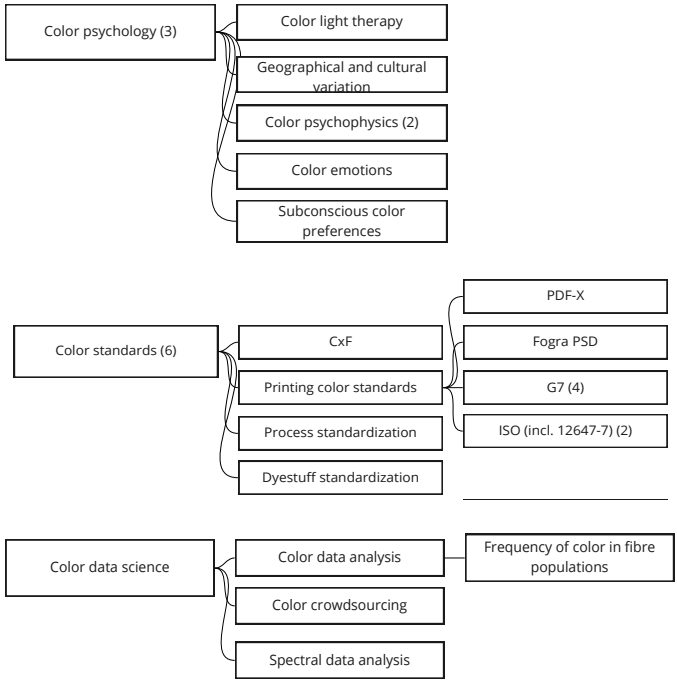